Research article	**Open Access**

# ATR-FTIR spectroscopy detects alterations induced by organotin(IV) carboxylates in MCF-7 cells at sub-cytotoxic/-genotoxic concentrations

Muhammad S Ahmad[1,2], Bushra Mirza[2], Mukhtiar Hussain[3], Muhammad Hanif[3], Saqib Ali[3], Michael J Walsh[1] and Francis L Martin*[1]

Address: [1]Lancaster Environment Centre, Lancaster University, Bailrigg, Lancaster LA1 4YQ, UK, [2]Department of Biochemistry, Faculty of Biological Sciences, Quaid-i-Azam University, Islamabad, Pakistan and [3]Department of Chemistry, Quaid-i-Azam University, Islamabad, Pakistan

Email: Muhammad S Ahmad - sheeraz_qau@yahoo.com; Bushra Mirza - bushra_dr@yahoo.com;
Mukhtiar Hussain - mukhtiar_chem@yahoo.co.uk; Muhammad Hanif - hanifns@yahoo.com; Saqib Ali - drsa54@yahoo.com;
Michael J Walsh - m.walsh@lancaster.ac.uk; Francis L Martin* - f.martin@lancaster.ac.uk

* Corresponding author

## Abstract

The environmental impact of metal complexes such as organotin(IV) compounds is of increasing concern. Genotoxic effects of organotin(IV) compounds (0.01 μg/ml, 0.1 μg/ml or 1.0 μg/ml) were measured using the alkaline single-cell gel electrophoresis (comet) assay to measure DNA single-strand breaks (SSBs) and the cytokinesis-block micronucleus (CBMN) assay to determine micronucleus formation. Biochemical-cell signatures were also ascertained using attenuated total reflection Fourier-transform infrared (ATR-FTIR) spectroscopy. In the comet assay, organotin(IV) carboxylates induced significantly-elevated levels of DNA SSBs. Elevated micronucleus-forming activities were also observed. Following interrogation using ATR-FTIR spectroscopy, infrared spectra in the biomolecular range (900 cm-1 – 1800 cm-1) derived from organotin-treated MCF-7 cells exhibited clear alterations in their biochemical-cell fingerprint compared to control-cell populations following exposures as low as 0.0001 μg/ml. Mono-, di- or tri-organotin(IV) carboxylates (0.1 μg/ml, 1.0 μg/ml or 10.0 μg/ml) were markedly cytotoxic as determined by the clonogenic assay following treatment of MCF-7 cells with ≥ 1.0 μg/ml. Our results demonstrate that ATR-FTIR spectroscopy can be applied to detect molecular alterations induced by organotin(IV) compounds at sub-cytotoxic and sub-genotoxic concentrations. This biophysical approach points to a novel means of assessing risk associated with environmental contaminants.

## Introduction

Organometallic compounds are agents that possess bonds between metal and carbon atoms [1]. The organometallic organotin(IV) compounds are characterized by the presence of at least one covalent C-Sn bond [2]. They contain a tetravalent {Sn} centre and are classified as mono-, di-, tri- or tetra-organotin(IV)s, depending on the number of alkyl (R) or aryl (Ar) moieties [2]. These compounds have been well-characterized structurally by infrared (IR), multinuclear NMR ($^{1}$H, $^{13}$C, $^{119}$Sn), $^{119m}$Sn Mossbauer spectroscopy and mass spectrometry [3,4]. Among the organotins, organotin carboxylates are important; these contain a Sn-O bond formed through the COO$^{-}$ group and exhibit a number of interesting structural features because of the tendency of the anionic group to coordinate inter- or intra-molecularly to tin(IV) [5].

Metal complexes, including organotin(IV) compounds, are used in a number of biomedical and commercial applications [6-9]. Their structural chemistry has attracted considerable attention owing to their anti-tumour activity [3,4]. Understanding their mechanism of action may lead to the development of new anti-tumour drugs as they are less toxic than platinum-based drugs [10]. Due to the diverse usage of these compounds, organic and inorganic forms of tin have accumulated in the food chain [10]. These compounds have varying degrees of toxicity, depending on the nature and number of alkyl groups bonded to the tin atom. Since a number of organotins are toxic [11], there is concern that widespread usage may result in adverse effects within environmental and biological systems. Exposure to di- or tri-methyl, butyl or phenyl tin induced aneuploidy in human lymphocyte cultures [12].

Taking into account the structural and biological diversity of organotin(IV) carboxylates [13-16], we set out to determine whether attenuated total reflection-Fourier-transform infrared (ATR-FTIR) spectroscopy might be a novel biophysical approach that would allow one to identify effects associated with typical environmental exposures to mono-, di- and tri-derivatives. In the oestrogen receptor-positive MCF-7 breast carcinoma cell line, cytotoxicity was ascertained using the clonogenic assay and, genotoxicity using the alkaline single-cell gel electrophoresis (comet) assay and the cytokinesis-block micronucleus (CBMN) assay. Although a cancer cell line, MCF-7 cells are robust cell model that have been used to examine genotoxic and toxic effects of candidate test agents [17-20]. Because they are metabolically proficient, they are also susceptible to possible metabolite effects [17,20]. Interrogation of IR spectral characteristics of cellular material previously exposed to organotin(IV) carboxylates might be a novel and non-destructive method for screening exposure effects at sub-cytotoxic and sub-genotoxic concentrations. The aim of this study was to determine whether IR spectra, in the biomolecular range (900 cm$^{-1}$ – 1800 cm$^{-1}$), derived from organotin-treated MCF-7 cells might exhibit molecular alterations compared to control cells following exposures as low as 0.0001 μg/ml; such an approach would highlight the general and potential applicability of mid-IR spectroscopy to signature toxic effects at sub-lethal concentrations of compounds. Such a novel biophysical approach might facilitate the determination of risk posed following typical environmental exposures to environmental contaminants.

## Methods

### *Chemicals*

Three groups of organotin(IV) carboxylates including mono-, di- and tri-derivatives of ligands were examined; these included: 3,4-methylenedioxy-6-nitrophenylpropenoic acid (**L1**), 3,4-methylenedioxyphenylpropenoic acid (**L2**) and 2,3-methylenedioxybenzoic acid (**L3**). The chemicals were newly synthesized and characterized by elemental analyses, IR spectra, multinuclear NMR ($^{1}$H, $^{13}$C, and $^{119}$Sn) and mass spectroscopy [14,15]. These test agents were previously investigated for their antibacterial, antifungal, anti-tumour and cytotoxic activities [14,15]. Solid test agents were dissolved in dimethylsulfoxide (DMSO) and subsequently added to culture media as solutions in DMSO (maximum concentration 1% v/v). All these compounds are crystalline solids, stable in air and soluble in common organic solvents; physical data is shown in Table 1.

**Table 1: Physical data for organotin(IV) derivatives of 3,4-methylenedioxy-6-nitrophenylpropenoic acid (L1), 3,4-methylenedioxyphenylpropenoic acid (L2) and 2,3-methylenedioxybenzoic acid (L3)**

| Compound no. | General Formula | Molecular Formula | Molecular Weight | M.P. (°C) | Yield (%) | Elemental Analysis % Calculated (Found) | | |
|---|---|---|---|---|---|---|---|---|
| | | | | | | C | H | N |
| **Ligand 1** | (L1) | $C_{10}H_{7}O_{6}N$ | 237 | 278 | 80 | 50.63 (50.45) | 2.95 (2.80) | 5.91 (5.87) |
| 1 | $Bu_3Sn(L1)$ | $C_{22}H_{33}O_{6}NSn$ | 527 | 143–146 | 89 | 50.09 (50.23) | 6.26 (6.30) | 2.65 (2.70) |
| 2 | $BuSnCl(L1)_2$ | $C_{24}H_{21}O_{12}N_{2}SnCl$ | 684.5 | 142–145 | 70 | 42.07 (42.24) | 3.07 (3.15) | 4.09 (4.04) |
| 3 | $Ph_3Sn(L1)$ | $C_{28}H_{21}O_{6}NSn$ | 587 | 207–210 | 80 | 57.24 (54.43) | 3.58 (3.60) | 2.38 (2.45) |
| 4 | $Bu_2Sn(L1)_2$ | $C_{28}H_{30}O_{12}N_{2}Sn$ | 706 | 196–199 | 73 | 47.60 (47.74) | 4.25 (4.16) | 3.97 (4.01) |
| **Ligand 2** | (L2) | $C_{10}H_{8}O_{4}$ | 192 | 238 | 90 | 62.50 (62.30) | 4.17 (4.23) | - |
| 5 | $Et_2Sn(L2)_2$ | $C_{24}H_{24}O_{8}Sn$ | 560 | 177–179 | 81 | 51.43 (51.39) | 4.28 (4.24) | - |
| 6 | $Me_2Sn(L2)_2$ | $C_{22}H_{20}O_{8}Sn$ | 532 | 247–248 | 73 | 49.62 (49.98) | 3.76 (3.70) | - |
| 7 | $Ph_3Sn(L2)$ | $C_{28}H_{22}O_{4}Sn$ | 542 | 180–181 | 84 | 61.99 (62.25) | 4.06 (3.98) | - |
| 8 | $Bu_2Sn(L2)_2$ | $C_{28}H_{32}O_{8}Sn$ | 616 | 134–136 | 68 | 54.54 (54.68) | 5.19 (5.08) | - |
| **Ligand 3** | (L3) | $C_{8}H_{6}O_{4}$ | 166 | 165 | 65 | 57.83 (57.72) | 3.61 (3.70) | - |
| 9 | $Bu_2Sn(L3)_2$ | $C_{24}H_{28}O_{8}Sn$ | 564 | 180–184 | 78 | 51.06 (51.20) | 4.96 (5.01) | - |
| 10 | $BuSn(L3)_3$ | $C_{28}H_{24}O_{12}Sn$ | 672 | 220–222 | 74 | 50.0 (50.10) | 3.57 (3.58) | - |
| 11 | $Me_3Sn(L3)$ | $C_{11}H_{14}O_{4}Sn$ | 330 | 150–151 | 68 | 40.0 (41.8) | 4.24 (4.01) | - |
| 12 | $Et_2Sn(L3)_2$ | $C_{20}H_{20}O_{8}Sn$ | 508 | 210–215 | 80 | 47.24 (47.23) | 3.94 (3.81) | - |

*Cell culture*

The human mammary carcinoma MCF-7 cell line was grown in complete medium comprising Dulbecco's modified essential medium (DMEM) supplemented with 10% heat-inactivated foetal calf serum, penicillin (100 U/ml) and streptomycin (100 μg/ml) [17,18]. Cells were grown in 5% $CO_2$ in air at 37°C in a humidified atmosphere and disaggregated using a trypsin (0.05%)/EDTA (0.02%) solution, to form single-cell suspensions prior to sub-culture or incorporation in experiments [18].

*The clonogenic assay*

Test agents were examined at three concentrations (0.1 μg/ml, 1.0 μg/ml and 10.0 μg/ml) with DMSO as vehicle control. Aliquots of $\approx 0.5 \times 10^3$ MCF-7 cells in 5 ml complete medium were exposed to the various treatments in T25 flasks [17,19,20]. Cells were exposed under incubation conditions of 5% $CO_2$ in air at 37°C in a humidified atmosphere for 24 h, after which medium was replaced with fresh medium [19]. Cells were then cultured for a further 7-day period, whereupon they were fixed with 70% ethanol (EtOH) and stained with 5% Giemsa; visible colonies were subsequently scored [17,19,20]. Results are presented as the mean ± SD of the average values derived from triplicate measurements for each treatment condition in three independent experiments.

*The alkaline single-cell gel electrophoresis (Comet) assay*

Microscope slides were cleaned with 70% EtOH and pre-coated on one side with a 1% solution of normal melting point (NMP) agarose in $H_2O$ [21]. MCF-7 cells grown to confluence in 12-well multi-well dishes were exposed in 1 ml of complete medium to various concentrations (0.01 μg/ml, 0.1 μg/ml and 1.0 μg/ml final concentrations) of test compound added as a solution in DMSO; DMSO was employed as the vehicle control and 1.0 μM benzo[*a*]pyrene (B[*a*]P) as positive control [21,22]. Subsequent steps were carried out under a darkened environment. After 2-h exposure, cells were disaggregated with trypsin/EDTA solution and re-suspended in 1 ml phosphate-buffered saline (PBS). To this, an equal volume of a warm ($\approx 37$°C) 1% solution of low melting point (LMP) agarose in PBS was added. A 150-μl aliquot containing $\approx 5 \times 10^4$ cells was added to coded slides. This LMP agarose layer was evenly spread through the application of a coverslip and allowed to set for 5 min on a cold surface. Coded slides were then submerged overnight in cold (4°C) lysis solution [buffer (2.5 M NaCl, 100 mM EDTA disodium salt, 10 mM Tris, 1% N-lauroyl sarcosine; adjusted to pH 10 with NaOH) with 1% Triton X-100 and 10% DMSO added prior to use] [22]. The slides were then transferred to a horizontal electrophoresis tank, covered in electrophoresis solution (0.3 M NaOH, 1 mM EDTA, pH > 13), and stored for 35 min at 10°C prior to electrophoresis at 0.8 V/cm and 300 mA for 35 min [21-23]. After electrophoresis, slides were neutralized with neutralization buffer (pH 7.5) and stained with ethidium bromide (20 ng/ml) after which comet-forming activity was scored by epifluorescence using a Leitz Dialux 20 EB microscope. Analyses for comet tail length (CTL; μm), % Tail DNA (% T DNA) and Olive tail moment (OTM) were performed using Comet IV software [23,24]. From duplicate slides, 100 images were analysed from individual experiments; each experiment was conducted independently twice.

### *The cytokinesis-block micronucleus (CBMN) assay*

MCF-7 cells were re-suspended in fresh complete medium and 3-ml aliquots ($\approx 1 \times 10^4$ cells) were seeded into Petri dishes containing 20 mm coverslips (Sarstedt, UK) [17]. After a 24-h incubation to allow attachment, cells were than exposed to test agents (0.01 μg/ml, 0.1 μg/ml and 1.0 μg/ml), as indicated; an equal volume of DMSO was employed as vehicle control and B[*a*]P (1.0 μM) as a positive control. Following a 24-h exposure, medium was replaced with fresh medium without treatment but containing 2 μg/ml cytochalasin-B [17,25]. The dishes were incubated for a further 24 h, whereupon cells were fixed with 70% EtOH and stained with 5% Giemsa. For each cell population examined, micronuclei in 1,000 binucleate MCF-7 cells, total number of micronuclei in 1,000 binucleate MCF-7 cells and the distribution of multiple micronuclei ($\leq 5$) in 1,000 binucleate MCF-7 cells was ascertained [17,25].

### *Attenuated total reflection-Fourier-transform infrared (ATR-FTIR) spectroscopy*

MCF-7 cells ($1/6^{th}$ of a confluent T75 flask in 6 ml complete medium) were cultured in 60-mm tissue culture dishes for 24 h prior to being treated with test compounds (0.0001 μg/ml, 0.01 μg/ml or 1.0 μg/ml); DMSO was employed as a vehicle control [19]. After 24-h exposure, cells were disaggregated with trypsin/EDTA solution, fixed in 70% EtOH, and applied to Low-E Microscope slides whereupon the suspensions were allowed to air-dry. Samples were then stored in a desiccated environment prior to analysis. IR spectra were acquired using a Bruker Vector 22 FTIR spectrometer with Helios ATR attachment that contained a diamond crystal (Bruker Optics, USA) [19]. Using a closed circuit television (CCTV) camera attached to the ATR crystal, ten random locations were interrogated across the samples [18]. Spectra were collected in ATR mode (8 cm$^{-1}$ spectral resolution, co-added for 32 scans) and were converted into absorbance using Bruker OPUS software. Sodium dodecyl sulfate (Sigma Chemical Co., UK) was used to clean the ATR crystal after analysis of each sample. Each time the crystal was cleaned, a new background reading was taken before recommencing spectral analysis. Spectra were baseline corrected and normalized to Amide I ($\approx 1650$ cm$^{-1}$) absorbance band using OPUS software [26]. IR spectra (900 cm$^{-1}$ – 1800 cm$^{-1}$) were analysed employing principal component analysis (PCA) using the Pirouette software package (Infometrix, USA) [27]. In PCA, each spectrum becomes a single point or score, in *n*-dimensional space and using selected principal components (PCs) as coordinates, the data was analysed for clustering when viewed in different directions. A 3-D scores plot on PCs selected to demonstrate best segregation of cell spectra from different treatment groups was obtained; derived loadings plots highlighted the wavenumbers contributing to variance [26,27].

## Results

### *The clonogenic assay*

The effects of organotin(IV) derivatives on the colony-forming ability of MCF-7 cells were ascertained (Figure 1). With all the test agents examined, dose-related decreases in colony formation were noted. Organotin(IV) derivatives of 3,4-methylenedioxy-6-nitrophenylpropenoic acid (**L1**) were the tri-organotin carboxylates $Bu_3Sn(L1)$ and $Ph_3Sn(L1)$ which were 100% toxic at $\geq 1.0$ μg/ml, the di-organotin $Bu_2Sn(L1)_2$ which was 100% toxic at $\geq 10.0$ μg/ml and the mono-organotin carboxylate $BuSnCl(L1)$ which was the least toxic in this series. The cytotoxicity of organotin(IV) derivatives of 3,4-methylenedioxyphenylpropenoic acid (**L2**) also increased in a dose-related fashion. The tri-organotin carboxylates $Ph_3Sn(L2)$ was 100% toxic at $\geq 1.0$ μg/ml whereas the di-organotin $Bu_2Sn(L2)_2$ was only toxic at $\geq 1.0$ μg/ml. Dose-related increases in the toxicities of organotin(IV) derivatives of 2,3-methylenedioxybenzoic acid (**L3**) were also observed (Figure 1). Following 24-h exposure to di-organotin carboxylates $Bu_2Sn(L3)_2$ or $Et_2Sn(L3)_2$, 100% cytotoxicity was noted $\geq 1.0$ μg/ml whereas the mono-organotin $BuSn(L3)_3$ was far less toxic. Surprisingly, in this series the least cytotoxic test agent was $Me_3Sn(L3)$. Each class of organotin(IV) derivative exhibited cytotoxicity following 24-h exposure in MCF-7 cells. In general, tri-organotins were more toxic than di-organotins with mono-derivative exhibiting the least toxicity in this assay.

### *Micronucleus-forming activity*

The micronucleus-forming activities of organotin(IV) carboxylates was assessed at three concentrations (Figure 2). Following 24-h exposure to all test agents, elevations in levels of micronuclei induced were observed. In the absence of cytotoxicity (Figure 1), micronucleus-forming activity was dose-related. Test agents such as $Bu_3Sn(L1)$, $Ph_3Sn(L1)$ and $Ph_3SnCl(L1)$ were all micronucleus-forming at the lowest concentration of 0.01 μg/ml, but higher concentrations were cytoxoxic. With a number of test agents including $Me_2Sn(L1)_2$, $Ph_3Sn(L2)$, $Et_2Sn(L3)_2$, $Bu_2Sn(L1)_2$, $Me_2Sn(L2)_2$, $Ph_3Sn(L2)$, $BuSn(L3)_2$ and $Et_2Sn(L3)_2$, a reversal in elevated micronucleus formation induced by lower exposures was noted in the presence of higher cytotoxic concentrations (Figure 2). Only the relatively non-cytotoxic organotin(IV) carboxylates including $BuSnCl(L1)$, $Et_2Sn(L2)_2$, $Me_2Sn(L2)_2$, $Bu_2Sn(L2)_2$ and $Me_3Sn(L3)$ exhibited micronucleus-forming activity throughout the concentration range tested; the least micronucleus-forming derivative was $BuSn(L3)_3$ (Figure 2). These results suggest that following 24-h exposure to $\geq 0.1$ μg/ml tri-organotins [$Bu_3Sn(L1)$, $Ph_3Sn(L1)$ or $Ph_3Sn(L2)$] except for $Me_3Sn(L3)$, cytotoxic effects were more important. Following exposure to di-organotins [$Bu_2Sn(L1)_2$, $Et_2Sn(L2)_2$, $Bu_2Sn(L3)_2$ or $Et_2Sn(L3)_2$] with the exception this time being $Bu_2Sn(L2)_2$, cytotoxicity was the more important endpoint with $\geq 1.0$ μg/ml. However, with the less cytotoxic mono-organotins [$BuSnCl(L1)_2$ and $BuSn(L3)_3$], elevated micronucleus formation was noted up to 10.0 μg/ml. In general, micronucleus-forming activity was of the order of mono- > di- > tri-organotin.

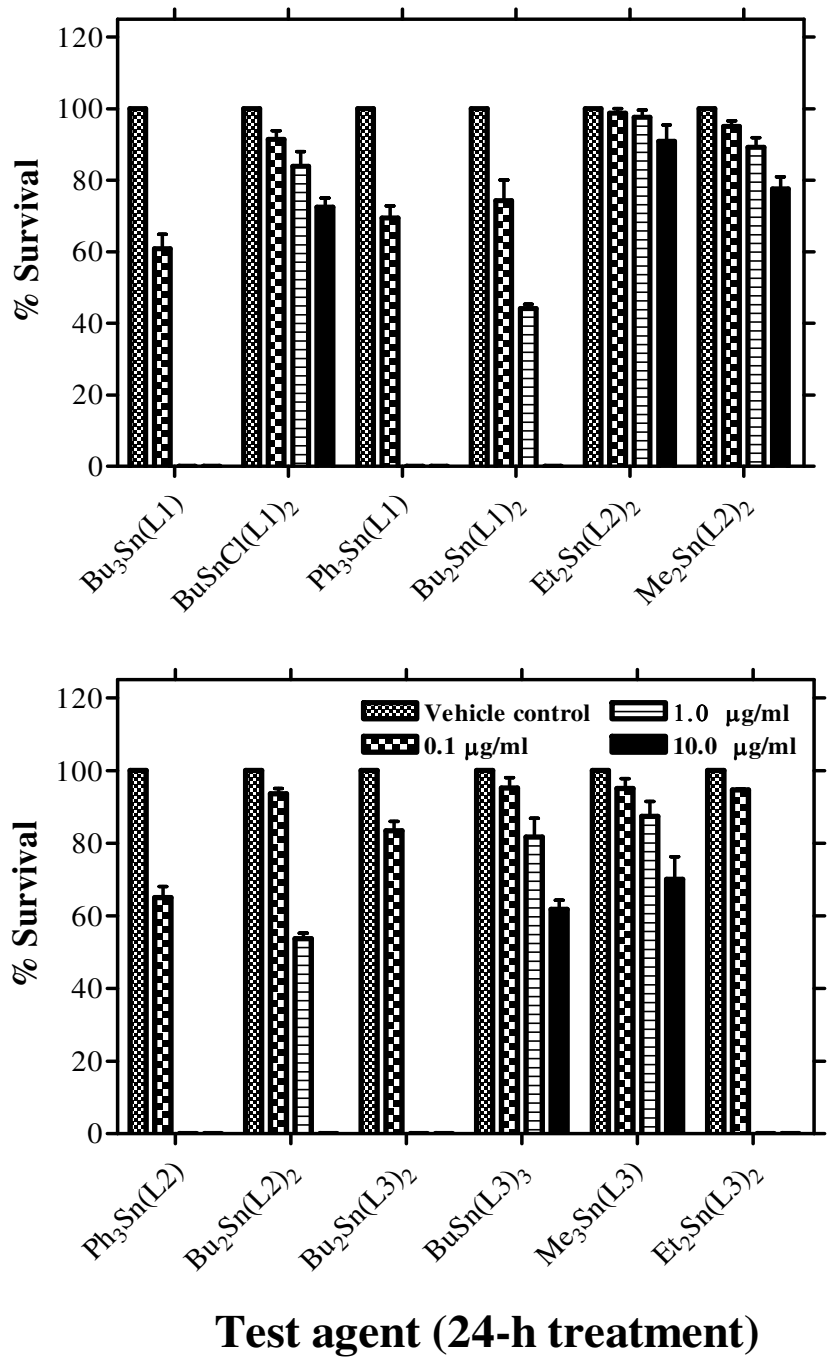

**Figure 1**
Effects of organotin(IV) carboxylates on per cent survival in MCF-7 cells. Cells (0.5 × $10^3$ cells) were seeded in 25 $cm^2$ flasks in the presence or absence of test agent and incubated for 24 h, as described in Materials and methods. Following addition of fresh medium, in the absence of exposure, cells were cultured undisturbed at 37°C and 5% $CO_2$ in a humidified atmosphere for 7 days. Surviving colonies were fixed and stained, and per cent survival was calculated by estimating the percentage of colonies counted over the number of cells initially seeded. Results are presented as the mean ± SD of the average values derived from triplicate measurements for each treatment condition in three independent experiments.

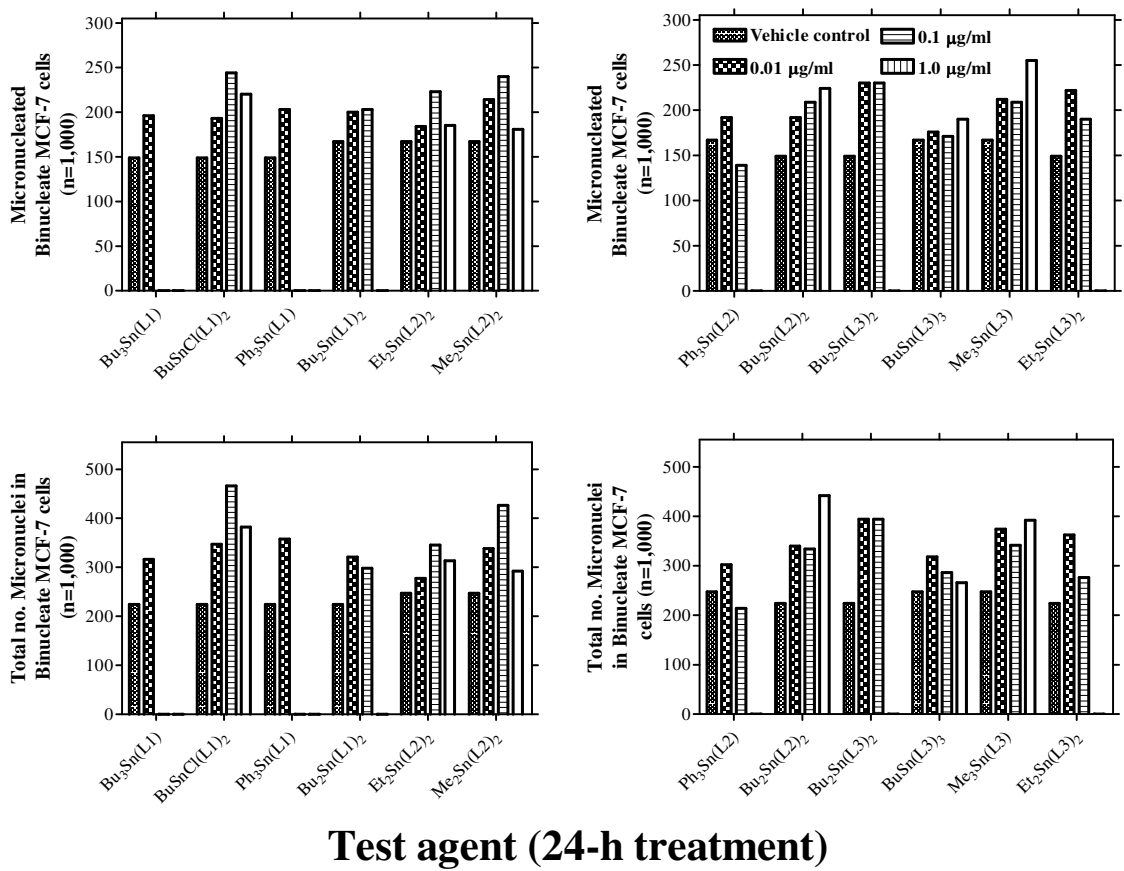

**Figure 2**
Micronucleus-forming activity of organotin(IV) carboxylates following 24-h exposure of MCF-7 cells. Cells were seeded as 3-ml aliquots ($\approx 1 \times 10^4$ cells) into 30 mm Petri dishes as described in Materials and methods. Following exposure the cells were blocked at cytokinesis by addition of fresh medium containing 2 μg cytochalasin B/ml. Cells were cultured for a further 24 h prior to fixation and staining with 5% Giemsa. Micronucleus formation was scored in 1,000 binucleate cells. In a corresponding control experiment, 1 μM benzo[*a*]pyrene induced a score of 255 micronucleated and 466 (the distribution being 135, 71, 28, 14, 7) total number of micronuclei in 1,000 binucleate MCF-7 cells. This compared to levels in vehicle control cells of 149 micronucleated and 219 (the distribution being 101, 30, 15, 2, 1) total number of micronuclei in 1,000 binucleate MCF-7 cells.

## *Comet-forming activity*

Except for instances where there was marked toxicity [*e.g.*, following 2-h exposure to 1.0 μg/ml Et$_2$Sn(L3)$_2$], all the organotin(IV) carboxylates were significantly ($P < 0.0001$) comet-forming as evidenced by an elevation of DNA SSBs determined by the alkaline comet assay (Table 2; Figure 3). Table 2 shows the comet-forming activities of organotin(IV) carboxylates derivatives in terms of CTL, %T DNA and OTM. MCF-7 cells were exposed for 2 h to either vehicle control (DMSO), three concentrations of test agent (0.01 μg/ml, 0.1 μg/ml or 1.0 μg/ml) or positive control (1.0 μM B[*a*]P). The results are a taken from a total of 400 randomly-selected nuclei from duplicate slides (*i.e.*, 100 nuclei scored per slide) and two independent experiments. Throughout all the experiments conducted, B[*a*]P was consistently and significantly (*P* < 0.0001) comet-forming. Although there were marked fluctuations in the background levels of DNA SSBs measurable in control MCF-7 cell populations, a remarkable consistency in the ability of all the organotin(IV) carboxylates to be comet-forming was noted (Figure 3; see Additional file 1). Following treatment with some organotin(IV) carboxylates [*e.g.*, Bu$_3$Sn (L1), Ph$_3$Sn(L1), Ph$_3$Sn(L2), Bu$_2$Sn(L3)$_2$], it may be that comet formation is being replaced by cytoxicity at the 1.0 μg/ml exposure level. However, in the presence of less toxic derivatives [*e.g.*, BuSnCl(L1)$_2$, Et$_2$Sn(L2)$_2$, Me$_3$Sn(L3)], Figure 3 does point to evidence of dose-related increases in formation of DNA SSBs. At all the concentrations tested, levels of DNA SSBs induced were not dissimilar to those observed following treatment with the positive control. However, with the alkaline comet assay it was more difficult to distinguish between the relative abilities of tri-organotin carboxylates, di-organotin carboxylates or mono-organotin carboxylates to induce elevations in DNA SSBs.

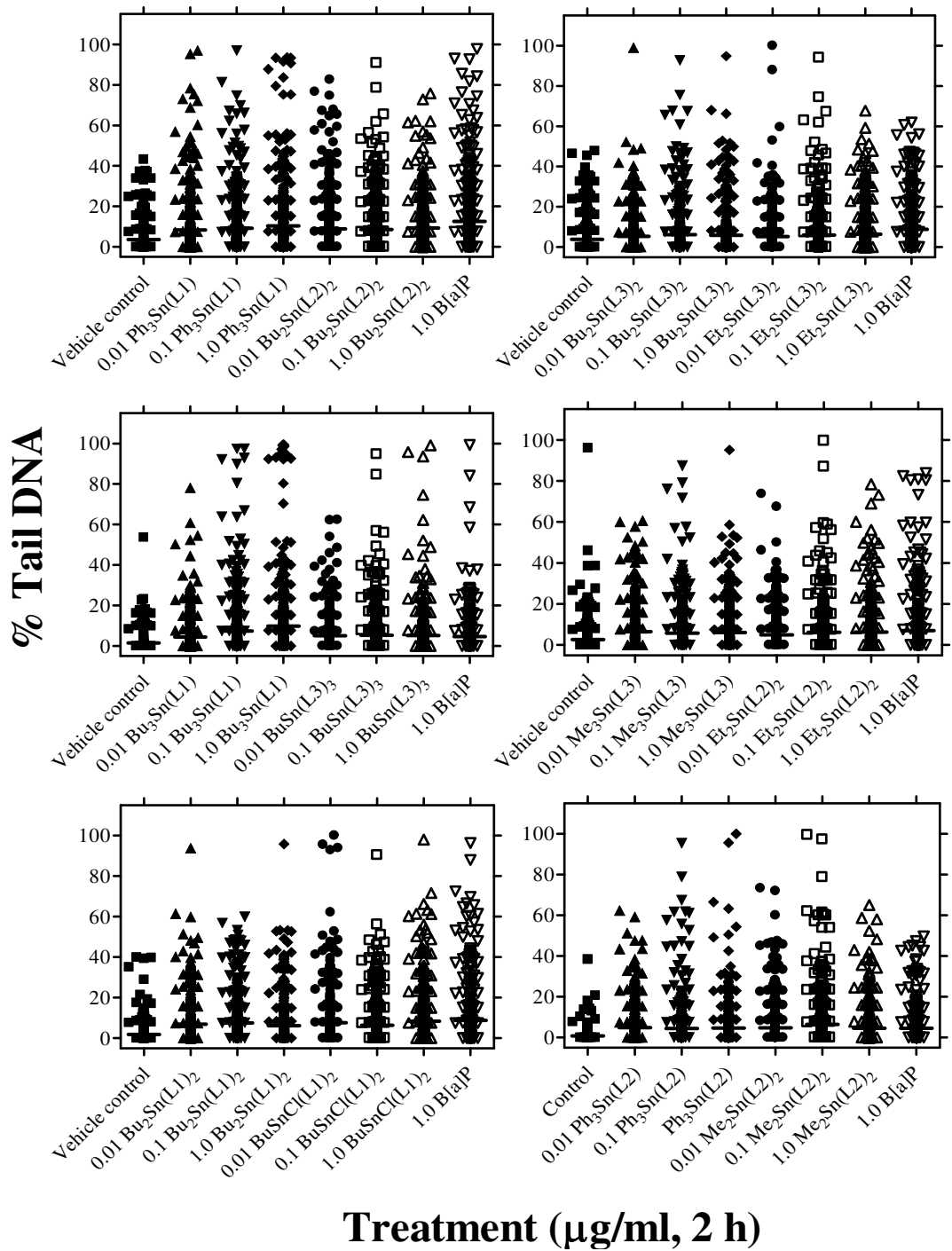

**Figure 3**
Comet-forming activity of organotin(IV) carboxylates in MCF-7 cells. Cells were grown to confluence in 12-well multi-well dishes and were exposed in 1 ml of complete medium to various concentrations (0.01 μg/ml, 0.1 μg/ml and 1.0 μg/ml final concentrations) of test agent added as a solution in DMSO; DMSO was employed as the vehicle control and 1.0 μM B[*a*]P as positive control. Following a 2-h exposure, cells were disaggregated with trypsin/EDTA prior to incorporation into the alkaline comet assay as described in the Materials and methods. As a measure of DNA damage, % Tail DNA (% T DNA) was used; for other measures of DNA damage including comet tail length (CTL; μm), and Olive tail moment (OTM), see Additional file 1.

**Table 2: Comet-forming activities of organotin(IV) derivatives**

| Compound | Comet Analysis | Mean ± 95% CI | | | | |
|---|---|---|---|---|---|---|
| | | Vehicle control | 0.01 µg/ml | 0.1 µg/ml | 1.0 µg/ml | B[a]P (1.0 µM) |
| **1 Bu$_3$Sn(L1)** | CTL (µm) | 38.7 ± 1.4 | 49.5 ± 2.0*** | 50.1 ± 1.7*** | 40.9 ± 1.8 | 47.6 ± 1.6*** |
| | % T DNA | 1.5 ± 0.4 | 4.4 ± 0.9*** | 7.4 ± 1.5*** | 9.9 ± 1.2*** | 4.7 ± 1.0*** |
| | OTM | 0.3 ± 0.1 | 1.3 ± 0.4*** | 2.1 ± 0.6*** | 2.0 ± 0.4*** | 1.4 ± 0.5*** |
| **2 BuSnCl(L1)$_2$** | CTL (µm) | 29.7 ± 0.9 | 39.5 ± 1.1*** | 40.7 ± 1.2*** | 38.3 ± 1.1*** | 44.7 ± 1.4*** |
| | % T DNA | 1.8 ± 0.5 | 7.6 ± 1.4*** | 6.5 ± 0.3*** | 8.5 ± 1.4*** | 8.9 ± 1.6*** |
| | OTM | 0.3 ± 0.1 | 1.5 ± 0.3*** | 1.4 ± 0.3*** | 1.6 ± 0.3*** | 2.1 ± 0.5*** |
| **3 Ph$_3$Sn(L1)** | CTL (µm) | 41.7 ± 2.3 | 52.9 ± 2.7*** | 51.1 ± 2.4*** | 48.9 ± 2.8*** | 56.8 ± 2.9*** |
| | % T DNA | 3.5 ± 0.6 | 8.3 ± 1.4*** | 9.3 ± 1.4*** | 10.4 ± 1.7*** | 12.68 ± 1.8*** |
| | OTM | 0.9 ± 0.2 | 2.8 ± 0.8*** | 2.7 ± 0.6*** | 3.5 ± 0.9*** | 4.0 ± 0.9*** |
| **4 Bu$_2$Sn(L1)$_2$** | CTL (µm) | 29.7 ± 0.9 | 41.4 ± 1.0*** | 40.65 ± 1.1*** | 38.0 ± 1.1*** | 44.7 ± 1.4*** |
| | % T DNA | 1.8 ± 0.5 | 7.0 ± 1.2*** | 7.55 ± 1.2*** | 6.2 ± 1.2*** | 8.9 ± 1.6*** |
| | OTM | 0.3 ± 0.1 | 1.4 ± 0.3*** | 1.49 ± 0.3*** | 1.2 ± 0.2*** | 2.1 ± 0.5*** |
| **5 Et$_2$Sn(L2)$_2$** | CTL (µm) | 33.7 ± 0.9 | 36.98 ± 1.1*** | 39.8 ± 1.3*** | 39.7 ± 1.1*** | 41.0 ± 1.5*** |
| | % T DNA | 2.7 ± 0.7 | 4.91 ± 0.99** | 6.2 ± 1.2*** | 6.3 ± 1.2*** | 7.1 ± 1.4*** |
| | OTM | 0.5 ± 0.1 | 1.06 ± 0.3*** | 1.4 ± 0.3*** | 1.3 ± 0.3*** | 1.9 ± 0.5*** |
| **6 Me$_2$Sn(L2)$_2$** | CTL (µm) | 30.2 ± 0.8 | 34.2 ± 1.1*** | 35.6 ± 1.2*** | 34.3 ± 1.0*** | 36.6 ± 1.2*** |
| | % T DNA | 0.8 ± 0.3 | 4.8 ± 1.1*** | 6.5 ± 1.2*** | 4.5 ± 1.1*** | 4.6 ± 0.9*** |
| | OTM | 1.1 ± 0.04 | 1.0 ± 0.3*** | 1.4 ± 0.5*** | 0.9 ± 0.2*** | 1.0 ± 0.2*** |
| **7 Ph$_3$Sn(L2)** | CTL (µm) | 30.2 ± 0.8 | 35.9 ± 1.1*** | 37.8 ± 1.2* | 32.4 ± 1.1* | 36.6 ± 1.15*** |
| | % T DNA | 0.8 ± 0.3 | 4.9 ± 1.1*** | 4.4 ± 1.2*** | 4.7 ± 1.1*** | 4.6 ± 0.9*** |
| | OTM | 1.1 ± 0.04 | 1.0 ± 0.2*** | 1.2 ± 0.5*** | 0.9 ± 0.2*** | 1.0 ± 0.9*** |
| **8 Bu$_2$Sn(L2)$_2$** | CTL (µm) | 41.7 ± 2.3 | 51.4 ± 2.4*** | 52.2 ± 2.4*** | 55.2 ± 2.9*** | 56.8 ± 3.0*** |
| | % T DNA | 3.5 ± 0.6 | 8.9 ± 1.4*** | 8.6 ± 1.3*** | 9.4 ± 1.2*** | 12.6 ± 1.8*** |
| | OTM | 0.9 ± 0.2 | 2.7 ± 0.6*** | 2.4 ± 0.5*** | 2.6 ± 0.5*** | 4.0 ± 0.9*** |
| **9 Bu$_2$Sn(L3)$_2$** | CTL (µm) | 37.5 ± 1.2 | 40.5 ± 1.3 | 39.4 ± 1.5 | 34.5 ± 1.3* | 50.0 ± 2.0*** |
| | % T DNA | 3.9 ± 0.85 | 5.3 ± 1.0 | 6.2 ± 1.1* | 5.8 ± 1.2* | 9.0 ± 1.7*** |
| | OTM | 0.8 ± 0.2 | 1.1 ± 0.4 | 1.5 ± 0.4* | 1.2 ± 0.31* | 3.2 ± 0.4*** |
| **10 BuSn(L3)$_3$** | CTL (µm) | 38.7 ± 1.4 | 45.8 ± 1.7*** | 45.2 ± 1.4*** | 44.8 ± 1.6*** | 47.6 ± 1.6*** |
| | % T DNA | 1.5 ± 0.4 | 5.0 ± 1.0*** | 5.2 ± 1.1*** | 5.2 ± 1.2*** | 4.7 ± 1.0*** |
| | OTM | 0.3 ± 0.1 | 1.3 ± 0.3*** | 1.3 ± 0.4*** | 1.4 ± 0.6*** | 1.4 ± 0.5*** |
| **11 Me$_3$Sn(L3)** | CTL (µm) | 33.7 ± 1.1 | 39.7 ± 1.3*** | 44.9 ± 1.3*** | 41.8 ± 1.2*** | 41.0 ± 1.5*** |
| | % T DNA | 2.7 ± 0.7 | 6.4 ± 1.2*** | 5.8 ± 1.2*** | 6.0 ± 0.1*** | 7.1 ± 0.4*** |
| | OTM | 0.5 ± 0.1 | 1.4 ± 0.3*** | 1.5 ± 0.4*** | 1.3 ± 0.3*** | 1.9 ± 0.5*** |
| **12 Et$_2$Sn(L3)$_2$** | CTL (µm) | 37.5 ± 1.2 | 42.1 ± 1.2*** | 40.8 ± 1.4** | 36.6 ± 1.4 | 50.0 ± 2.0*** |
| | % T DNA | 3.9 ± 0.85 | 5.2 ± 1.07 | 5.9 ± 1.2* | 6.4 ± 1.16** | 9.0 ± 1.7*** |
| | OTM | 0.8 ± 0.2 | 1.3 ± 0.3 | 1.4 ± 0.4* | 1.4 ± 0.3** | 3.2 ± 0.4*** |

\*\*\* *P* < 0.0001; \*\* *P* < 0.001; \* *P* < 0.01

### *A biochemical-cell fingerprint*

The effects of 24-h exposure of MCF-7 cells with 0.0001 μg/ml Bu$_3$Sn(L1) (a mono-organotin carboxylate) (Figure 4), Ph$_3$Sn(L2) (a di-organotin carboxylate) (Figure 5) or Me$_3$Sn(L3) (a tri-organotin carboxylate) (Figure 6) on the consequent IR spectrum of 70% EtOH-fixed cellular material, compared to vehicle control are shown. To see effects induced by all the organotin(IV) carboxylates tested at all exposures (0.0001 μg/ml, 0.01 μg/ml or 1.0 μg/ml), see Additional file 1. Following fixation, MCF-7 cells were applied to Low-E reflective glass slides and interrogated using the ≈ 250 μm × 250 μm octagon-shaped sampling area employed in this study for ATR-FTIR spectroscopy [18]. An IR spectrum of cellular biochemistry was obtainable from such 70% EtOH-fixed cells [19]. In the biomolecular region (900 cm$^{-1}$ – 1800 cm$^{-1}$), clear induced alterations in the "biochemical-cell fingerprint" were associated with organotin-treated cells (Figures 4, 5, 6; see Additional file 1). In this study, variability was noted in control spectra between 1400 cm$^{-1}$ and 1500 cm$^{-1}$; however, in order to distinguish intra-class differences (*i.e.*, vehicle control *vs.* treatment) one needs to look at the direction of the PCs. The most marked differences were noted in the spectral region ≈ 1300 cm$^{-1}$ – 1500 cm$^{-1}$; these would be associated with amide III absorptions (predominantly C-N stretching) with significant contributions from CH$_2$ stretching vibrations of carbohydrate residues (≈ 1280 cm$^{-1}$ – 1360 cm$^{-1}$), but some of these wavenumbers would be associated with intra-class as opposed to inter-class variance (Figures 4, 5, 6). However, in Figure 5 PC1 mostly excludes this aforementioned region as a contributor to variance but highlights 1225 cm$^{-1}$ (asymmetric phosphate), which consistently contributed to variance. Additionally, in some cases [Bu$_2$Sn(L1)$_2$, Et$_2$Sn(L2)$_2$, Me$_2$Sn(L2)$_2$, BuSn(L3)$_3$ and Me$_3$Sn(L3)], glycogen (≈ 1030 cm$^{-1}$) was observed as an important contributing factor. Of note was the fact that except for BuSnCl(L1)$_2$, separation of IR spectra derived from 0.0001 μg/ml exposed and control MCF-7 cells was readily achievable along PCs 1, 2 and 3 (see Additional file 1). This suggested that organotin(IV) carboxylates (tri-, di- or mono-derivatives) are capable of inducing profound biomolecular alterations even at low concentrations.

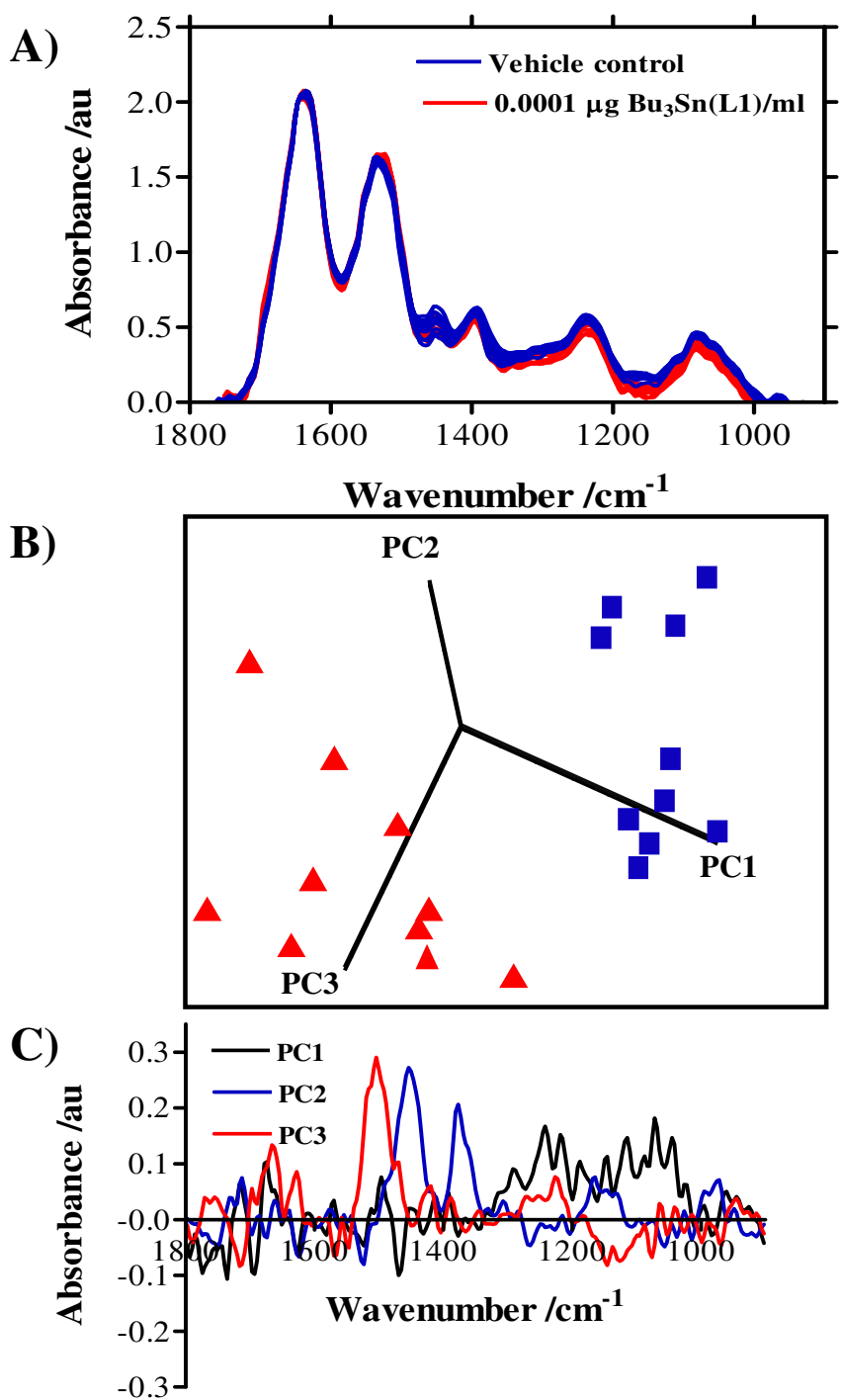

**Figure 4**
Average IR spectra of MCF-7 cells with corresponding scores plots plotted on PCs 1, 2 and 3 and loadings plots following 24-h exposure to 0.0001 μg $Bu_3Sn(L1)$/ml. Cells were seeded into 60-mm Petri dishes and allowed to attach for 24 h prior to exposure. Following exposure, cells were disaggregated with trypsin/EDTA solution, fixed in 70% EtOH, and applied to Low-E Microscope slides whereupon the suspensions were allowed to air-dry. **A)** 10 IR spectra were derived (vehicle control in blue, treated in red). **B)** From subsequent multivariate analysis, a 3-D scores plot on PCs selected to demonstrate best segregation of vehicle control (blue) *vs.* exposure (red) was constructed. **C)** Following cluster analysis, loadings plots were constructed on each relevant PC to identify the wavenumbers most responsible for segregation of clusters.

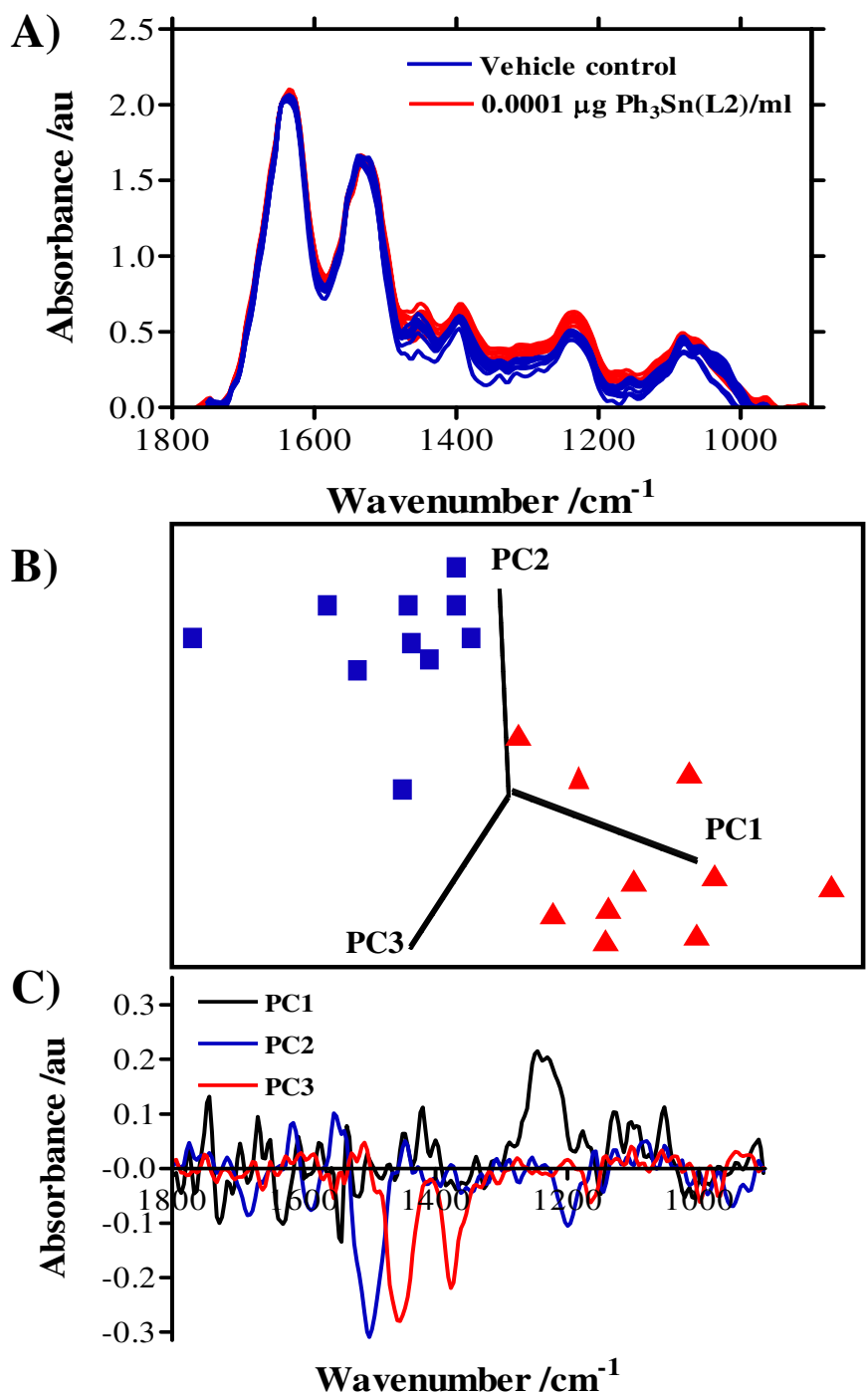

### Figure 5
Average IR spectra of MCF-7 cells with corresponding scores plots plotted on PCs 1, 2 and 3 and loadings plots following 24-h exposure to 0.0001 μg Ph$_3$Sn(L2)/ml. Cells were seeded into 60-mm Petri dishes and allowed to attach for 24 h prior to exposure. Following exposure, cells were disaggregated with trypsin/EDTA solution, fixed in 70% EtOH, and applied to Low-E Microscope slides whereupon the suspensions were allowed to air-dry. **A)** 10 IR spectra were derived (vehicle control in blue, treated in red). **B)** From subsequent multivariate analysis, a 3-D scores plot on PCs selected to demonstrate best segregation of vehicle control (blue) *vs.* exposure (red) was constructed. **C)** Following cluster analysis, loadings plots were constructed on each relevant PC to identify the wavenumbers most responsible for segregation of clusters.

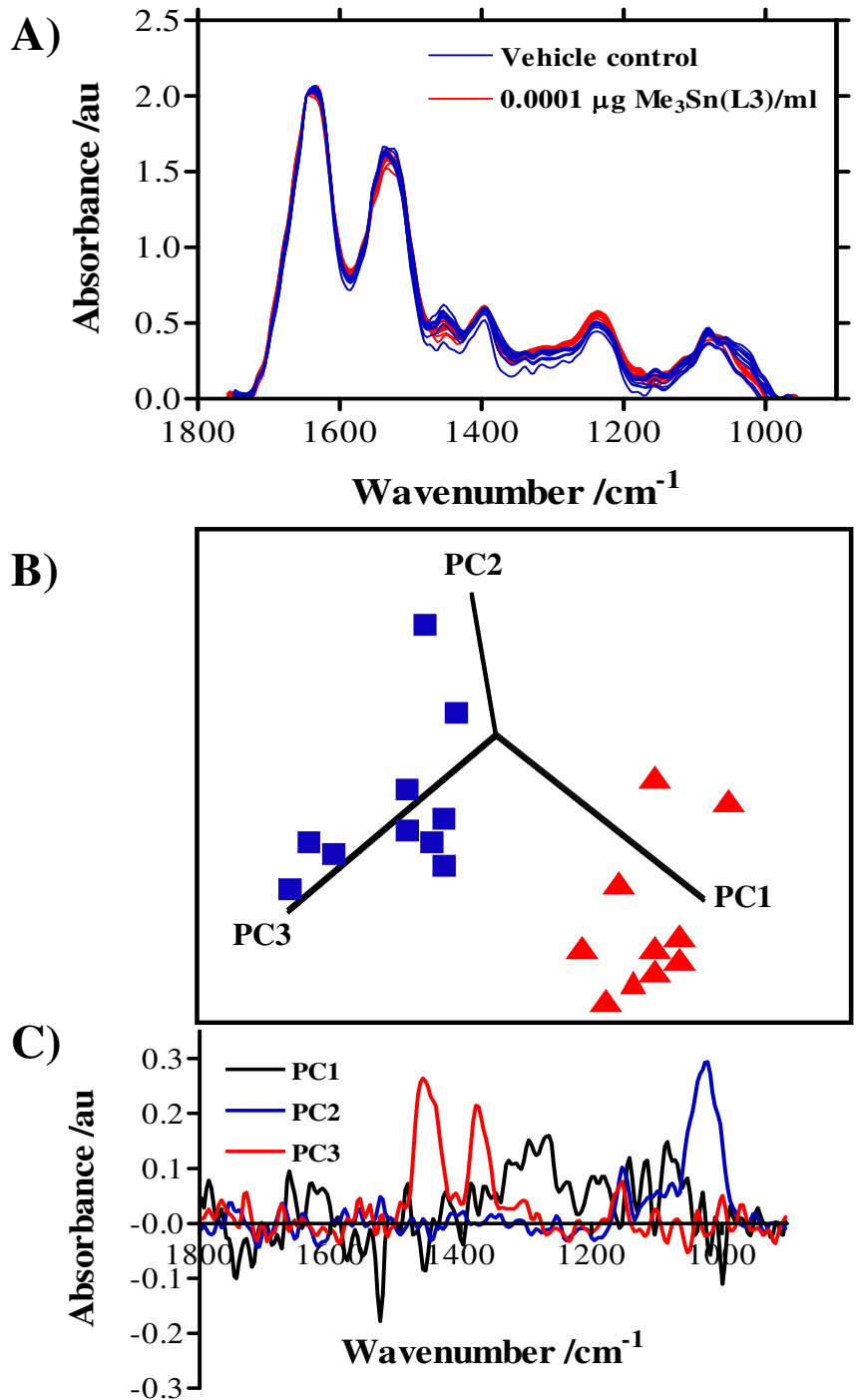

**Figure 6**
Average IR spectra of MCF-7 cells with corresponding scores plots plotted on PCs 1,2 and 3 and loadings plots following 24-h exposure to 0.0001 μg Me$_3$Sn(L3)/ml. Cells were seeded into 60-mm Petri dishes and allowed to attach for 24 h prior to exposure. Following exposure, cells were disaggregated with trypsin/EDTA solution, fixed in 70% EtOH, and applied to Low-E Microscope slides whereupon the suspensions were allowed to air-dry. **A)** 10 IR spectra were derived (vehicle control in blue, treated in red). **B)** From subsequent multivariate analysis, a 3-D scores plot on PCs selected to demonstrate best segregation of vehicle control (blue) *vs.* exposure (red) was constructed. **C)** Following cluster analysis, loadings plots were constructed on each relevant PC to identify the wavenumbers most responsible for segregation of clusters.

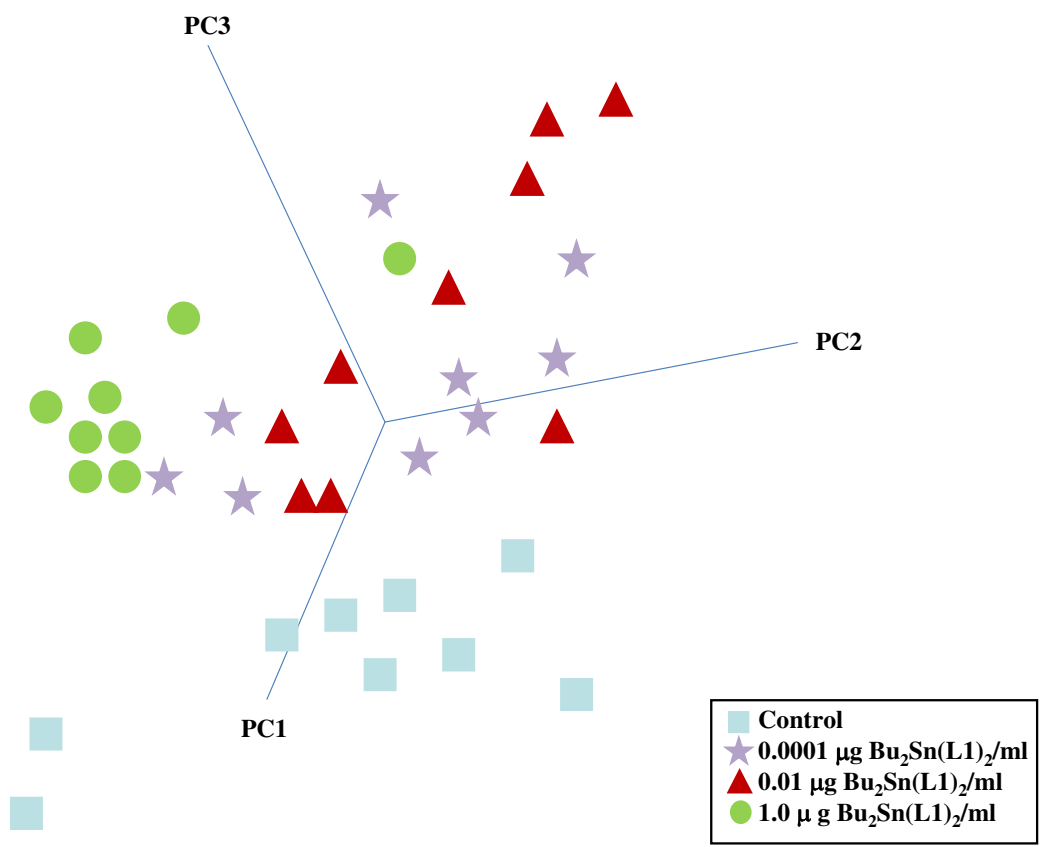

**Figure 7**
3-D scores plot on PCs selected to demonstrate best segregation of MCF-7 cell IR spectra derived following 24-h exposure to $Bu_2Sn(L1)_2$, as indicated. IR spectra were collected using ATR-FTIR spectroscopy. Each spectrum was expressed in terms of chosen PCs using Pirouette software and rotated to identify segregation of different clusters. Each symbol represents a single spectrum as a single point in "hyperspace". The 3-D scores plot represents the segregation of cell populations following different exposures (*i.e.*, vehicle control *vs.* 0.0001 μg/ml *vs.* 0.01 μg/ml *vs.* 1.0 μg/ml organotin(IV) carboxylate) on PC1, PC2 and PC3.

　　PCA is employed as a multivariate technique to reduce the number of correlated variables (*i.e.*, IR spectra) to a smaller number of independent variables, known as PCs that account for the majority of the variance of the original data; this thereby reduces its dimensionality. Taking IR spectra derived from cellular material previously exposed to $Bu_2Sn(L1)_2$, it was notable to see that one could cluster the dose-related effects of this test agent (Figure 7). A scores plot along PCs 1, 2 and 3 was constructed and rotation allowed selection of the view that highlighted best segregation (*i.e.*, vehicle control *vs.* 0.0001 μg/ml *vs.* 0.01 μg/ml *vs.* 1.0 μg/ml organotin(IV) carboxylate). In such a scores plot, nearness in space implies similarity whilst distance points to segregation in the derived IR spectral fingerprint of the interrogated cellular material. PC1 is defined in the direction of maximum variance of the whole data set whilst PC2 is the direction that describes the maximum variance in the orthogonal subspace to PC1. Subsequent PCs are taken orthogonal to those previously chosen and describe the maximum of the remaining variance. Generally, only the first few PCs are required to describe most of the information contained in the original data set. Figure 7 strongly supports the notion that ATR-FTIR spectroscopy combined with multivariate analysis is a powerful new approach for identifying exposure effects in target cell populations.

**Discussion**

The acute effects of organotin(IV) compounds, used as biocides against a number of organisms, have been extensively investigated [2]. Their mechanism of toxicity remains obscure. It is assumed that these compounds are capable of reacting with cell membranes leading to leakage, accelerating ion exchange processes, and inhibiting oxidative or photochemical phosphorylation. Among $R_nSnX_{(4-n)}$ compounds, the most toxic appear to be the $R_3SnX$ [2,28,29]. In this study, we set out to apply ATR-FTIR spectroscopy in order to ascertain whether IR spectra, in the biomolecular range (900 cm$^{-1}$ – 1800 cm$^{-1}$), derived from organotin-treated MCF-7 cells might exhibit molecular alterations compared to control cells following exposures to sub-cytotoxic and sub-genotoxic concentrations of test agent. Effects of a panel of mono-, di- or tri-organotin(IV) carboxylates in MCF-7 cells were examined. Cytotoxicity was assessed in the clonogenic assay whilst genotoxicity was determined using the CBMN and alkaline comet assays.

In the 0.01 μg/ml to 10.0 μg/ml concentration range, all the organotin(IV) carboxylates tested induced clear dose-related increases in cytotoxicity (Figure 1). However, with six of the derivatives [$BuSnCl(L1)_2$, $Et_2Sn(L2)_2$, $Me_2Sn(L2)_2$, $BuSn(L3)_3$ and $Me_3Sn(L3)$] >60% survival was still observed following 24-h exposure to 10.0 μg/ml. In line with these observations, we examined the ability of the test agents to induce elevated levels of micronuclei or DNA SSBs (Figures 2 and 3). Even at the lowest concentration of 0.01 μg/ml, marked elevations in micronucleus formation [except for $BuSn(L3)_3$] and significant ($P < 0.0001$) increases in DNA SSBs [except for $Bu_2Sn(L3)_2$] were induced following either 24-h or 2-h exposures, respectively. At higher concentrations of test agent, apparent reversals in the induction of genotoxic effects were probably associated with cytotoxicity. An example of this was the dose related effects of $BuSnCl(L1)_2$; in the CBMN assay (Figure 2) and the alkaline comet assay there are elevations in genotoxicity with increasing concentration up to 0.1 μg/ml. However, at the higher 1.0 μg/ml concentration there is an apparent falling off in the levels of micronucleus formation (Figure 2) or DNA SSBs (Table 2). Although not as marked as with other organotin(IV) carboxylates, there is a measurable cytotoxic effect in MCF-7 cells associated with this concentration of test agent.

One might surmise that environmental effects of such agents might be a consequence of exposures an order-of-magnitude or more lower than those routinely used in short-term (geno)toxicity assays. The requirement for an appropriate concentration range in order to detect measurable effects in such assays is a major limitation as it is then difficult to make meaningful extrapola-tions to lower but more environmentally-relevant concentrations. To this end, we applied ATR-FTIR spectroscopy to ascertain whether IR spectral changes might be identifiable in cells exposed to concentrations as low as 0.0001 µg/ml (Figures 4, 5, 6). Following IR spectral interrogation of treated *vs.* vehicle control cells, PCA was used to derive scores plots along PCs 1, 2 and 3. Except for two of the test agents [BuSnCl(L1)$_2$ and Bu$_2$Sn(L2)$_2$], marked separation of vehicle control *vs.* treatment clusters was observed (see Additional file 1). In the loadings plots along the PCs used, there was a consistency in the identification of peaks in the spectral region ≈ 1300 cm$^{-1}$ – 1500 cm$^{-1}$ as being the predominant contributors to variance segregating the clusters. These loadings would be associated with amide III absorptions (predominantly C-N stretching) with significant contributions from CH$_2$ stretching vibrations of carbohydrate residues (≈ 1280 cm$^{-1}$ – 1360 cm$^{-1}$) (Figures 4, 5, 6). However, PC1 often seemed most responsible for inter-class variance and pointed to wavenumbers primarily associated with DNA/RNA conformational changes *e.g.*, 1225 cm$^{-1}$. Identifying wavenumbers responsible for inter-class variance would be important as these might give a signature of low-dose toxic effects associated with organotin(IV) carboxylates [19].

Di-n-butyltin and tri-n-butyltin chloride are known to induce apoptosis *in vitro* in rat thymocytes [30,31]. The apoptotic pathway induced starts with an increased influx of Ca$^{2+}$ ions and is followed by the release of the cytochrome c from mitocondria, activation of caspases and finally DNA fragmentation [31]. Diethyltindichloro(1,10-phenanthroline) inhibits cancer cell growth and also alters the surface of the cancer cell membrane [32,33]. The binding efficiency to DNA of organotin compounds depends on the coordination number and nature of groups bonded to the central tin atom [32]. The phosphate group of the DNA sugar backbone usually acts as an anchoring site and often results in the stabilization of the tin centre as an octahedral stable species [32,33].

Organotin(IV) compounds may have applications in cancer chemotherapy [32,34-38]. Recently, thirty interesting inorganic pharmaceuticals, four of which are tin compounds [32,33,39] have been described. If such agents have significant usage, it will be important to be able to identify measurable effects following low-level exposures to such potentially highly toxic derivatives.

## Conclusion

Ascertaining the biological effects of environmental contaminants, especially if their cell or intracellular targets are unknown, remains a difficult challenge. Conventional cell-biology assays generally measure a single end-effect (*e.g.*, cytotoxicity, micronucleus formation or comet-forming activity). Concentrations much higher than those generally found in the environment are usually required to induce such end-effects and one needs to extrapolate from such experimental settings. However, our study demonstrates that ATR-FTIR spectroscopy can be used to identify signature biological effects at concentrations much lower than those required to induce cytotoxicity or genotoxicity. Application of this biophysical approach is a novel means of assessing risk associated with environmental contaminants. Whether the induced IR spectral alterations observed are a signature of exposure to organotin(IV) carboxylates or a more general molecular alteration in response to environmental contaminants will be examined by comparing the biochemical-cell fingerprints of different classes of environmental contaminants.

## Additional material

**Additional file 1**
*SupplementaryData-PMC Biophysics Ahmad MS et al. Additional data to Ahmad* et al. *(2008) containing organotin(IV) carboxylates-specific effects at increasing concentrations in MCF-7 cells.*
Click here for file
[http://www.biomedcentral.com/content/supplementary/1757-5036-1-3-S1.pdf]

## Acknowledgements
This study was supported by "Indigenous PhD scholarship 5000 Program" and "International Research Support Initiative Program" of Higher Education Commission of Pakistan. Research in F.L.M.'s laboratory is supported by Rosemere Cancer Foundation.